\documentclass[manuscript,aps]{revtex4}
\usepackage{graphicx}

\begin{document}


\title{Gravity-like potential traps light and stretches supercontinuum in
photonic crystal fibers}

\author{A.V. Gorbach and D.V. Skryabin$^{*}$}

\affiliation{Centre for Photonics and Photonic Materials,
Department of Physics, University of Bath, Bath BA2 7AY, UK\\
$^*$e-mail: d.v.skryabin@bath.ac.uk}


\maketitle
\newpage

{\bf The use of  photonic crystal fibers pumped by  femtosecond
pulses has enabled the generation of   broad optical supercontinua
with nano-joule input energies \cite{ranka,dudley}. This striking
discovery has  applications ranging from spectroscopy and
metrology \cite{prl} to telecommunication \cite{tur} and medicine
\cite{bio}. Amongst the physical principles underlying
supercontinuum generation are soliton
fission \cite{gus}, a variety
of four-wave mixing processes \cite{pre,gorbach,wil,efim},
Raman induced soliton self-frequency shift
\cite{aus,science}, and dispersive wave generation  mediated by
solitons \cite{gus,science,italy}. Although all of the above effects
contribute to supercontinuum generation none of
them can explain the generation of  blue and violet light
from  infrared femtosecond pump pulses, which has been seen  already in the first
observations of the supercontinuum in photonic crystal fibers \cite{ranka}.
In this work we argue that the most
profound role in the shaping of the short-wavelength edge of the
continuum is played by the effect of radiation trapping in a
gravity-like potential created by accelerating solitons. The
underlying physics of this effect has a straightforward analogy with
the  inertial forces acting on an observer moving with a constant  acceleration.}

A common method of producing  broad optical spectra from a
spectrally narrow femtosecond pump relies upon the fact that
fibers with a silica core (having few micron diameters)
and surrounded by a photonic crystal cladding with various geometries,
can be designed to have a zero group velocity dispersion (GVD or simply dispersion)
wavelength $\lambda_0$ in a proximity of $800$nm
\cite{ranka,dudley,gus,gorbach}, which matches the wavelength of a mode-locked
Ti-Sapphire laser. Close to $\lambda_0$ the dispersion
is small and so  the input pulse can sustain a high peak
power over a considerable length, which together with the small size of the fiber core,
enhances variety of nonlinear effects, thus initiating a
dramatic spectral broadening known as supercontinuum generation
\cite{ranka,dudley,gus,pre,gorbach,wil,efim,italy}. The
measurements of supercontinuum spectra generated with a
femtosecond pump have been successfully reproduced in numerical
modeling using the generalized nonlinear Schr\"odinger equation (see
Eq. (2) in \cite{dudley}).
A typical example of the spectral evolution calculated using this model and
leading to the supercontinuum generation by a 200fs pulse
is shown in Figs. 1(a,b). The dispersion of the modeled fiber is shown in Fig. 1(c).
Having $\lambda_0$ close to $800$nm shifts the range of the anomalous GVD
towards much shorter wavelengths than in  telecom fibers and
extends the range in which  optical solitons can exist.

The first stage ($0<z<0.1$m) of the spectral broadening in
Fig. 1(a) is due to the formation of symmetric spectral sidebands through
the well known effect of self-phase modulation, which is  mediated by
instantaneous Kerr nonlinearity \cite{spm,agrawal}. The slope of the
group velocity dispersion and its sign change do not play a
significant role during this stage. At the second stage of the
spectral evolution ($0.1$m$<z<0.2$m) the asymmetry between the
short- and long-wavelength   edges of the spectrum becomes
pronounced.  The long-wavelength edge is shaped by the process of
soliton formation (soliton fission) \cite{gus}. With further propagation
the solitons are further
shifted towards  longer wavelengths  by  intrapulse Raman scattering
\cite{agrawal,aus,science,bel}. If dispersion is anomalous then
the longer wavelengths correspond to  smaller group velocities,
and so the solitons continuously slow down with propagation.
The spectrum at the short-wavelength edge is created by  resonant
radiation from the  solitons \cite{gus,italy} and by the four-wave
mixing of  solitons with  dispersive waves \cite{pre,gorbach}.
The dispersive waves generated by both of these mechanisms initially have group
velocities smaller  than the ones  of the solitons and lag behind the latter
\cite{gorbach}. However, the solitons  are continuously
decelerated and therefore the dispersive waves catch up with and start
to interact with the solitons. This interaction leads to  emission of new frequencies at even
shorter wavelengths \cite{gorbach}.

Importantly, the wave packets  emitted through the above
process experience normal group velocity dispersion, see Figs.
1(a,c). Normal dispersion can not be compensated for by the focusing
nonlinearity of the silica and hence these packets, however strong
or weak,  are expected to spread out in time. Therefore the
efficiency of their interaction with the soliton (which is
proportional to the  amplitude of the wave packet \cite{pre}) should
be noticeably reduced after  distances comparable to the dispersion
length ($\sim 10$cm) and the associated spectral broadening should cease to continue.
However,  experimental and numerical
observations demonstrate continuous frequency shifting of the short-wavelength edge
of the continuum without significant energy loss, see Fig. 1(a).
Also, simultaneous spectral and time-domain analysis of the experimental
and numerical data by means of the cross-correlated frequency
resolved optical gating (XFROG)  gives a full impression of the formation
of  soliton like structures not only at
the long-wavelength edge of the spectrum, where dispersion  is anomalous, but at
its short-wavelength edge as well, where dispersion is normal and usual bright
solitons can not exist, see Fig. 1(b) and Refs. \cite{gorbach,josab,fins}.
One of the properties of the short-wavelength pulses
is that their wavelength gets continuously
shorter  at roughly the same rate at which  the soliton wavelength gets longer.
One can also see from the
Fig. 1(b), that the solitons and the pulses at the short-wavelength edge are paired
together, so that every soliton has an associated non-dispersive
pulse on the opposite side of the spectrum.

Even without  supercontinuum generation one can simply take two appropriately delayed
and group velocity matched pulses with spectra across the zero GVD wavelength and
observe their bound propagation,  in which the dispersive spreading of
the pulse in the normal GVD range been canceled and its frequency is being
constantly blue shifted, see Figs. 2(a-c).  This effect
has been reported in the series of papers by Nishizawa and Goto \cite{goto}.
Their latter work and some of the papers reporting the same effect in the
context of  supercontinuum generation \cite{josab,fins} have put forward the
arguments that the blue shift of the pulse in the normal GVD range is caused by
cross-phase modulation (XPM) mediated by  Kerr nonlinearity \cite{agrawal}. However,
the origin of the suppression of  dispersive spreading, which is crucial for the
continuous blue shift,  has remained unclear. In what follows we demonstrate
that the above supercontinuum expansion and the Nishizawa-Goto experiments
are explained by the existence of a new kind of
two-frequency soliton-like states. Compensation of the XPM of the short-wavelength component
is one of the necessary conditions for the existence of these soliton states. This
rules out XPM as a possible reason for the continuous spectral shift towards shorter wavelengths.
It is instructive to remove the Raman effect from the numerical model, in which case immediate
dispersive spreading of the short-wavelength component and cancelation of the frequency shifts occur,
see Figs. 2(d-e). The only interaction between the
two pulses remaining under these circumstances is due to XPM \cite{agrawal}.
It is thus clear that the latter, at least in its own right,
does not provide the conditions necessary for trapping
and frequency conversion of the short-wavelength radiation.

The two pulse experiments and modelling, see \cite{goto} and Fig. 2,
suggest that to describe continuous broadening of the supercontinuum  it is sufficient
to focus on the coupling between its spectral edges and neglect the middle part of the spectrum.
By analyzing the numerical data one can find  that  the short-wavelength pulse
is typically much weaker than the associated long-wavelength soliton.
Therefore, only the nonlinear refractive index change induced by the soliton
is significant and ought to be accounted for. The
equation for the slowly varying amplitude $A$ of the short-wavelength pulse, which is
readily derived from the generalized nonlinear Schr\"odinger equation,
takes the form
\begin{equation}
-i\partial_z A-(i/v_s)\partial_t A=[d\partial^2_t +U] A,\label{cnls}
\end{equation}
where $U=2\gamma P -\tau_R\gamma\partial_{t}P$.
The  electric field of the pulse
is $Ae^{ik_0 z-i\omega_0 t}+c.c.$, where
$\omega_0$ is the central frequency of the pulse and
$k_0$ is the corresponding propagation constant.
The notations here and below are as follows:
$z$ is the coordinate along the fiber, $t$ is time,
$v_{s,l}$ are the group velocities
of the short- and long-wavelength pulses, respectively,
and $d=\lambda^2D/(4\pi c)$ is  the dispersion
coefficient (see Fig. 1(c) for the $D(\lambda)$ plot),
$P$ is the soliton  power, $\gamma\approx 0.02$1/W/m
is the nonlinear parameter  \cite{gorbach},
$\tau_R\approx 1.46$fs is the Raman time \cite{agrawal}.
The time dependence of the soliton power  in the laboratory frame of reference is
$P=4 P_0/(e^{\tau/w}+e^{-\tau/w})^2$, where $w$ is the soliton width,
$\tau=t-z/v_l+gz^2/2$ and $g=8\tau_R\gamma^2 P^2_0/15$
is the soliton acceleration  \cite{bel}.

Since the Kerr nonlinearity in optical fibers is focusing the soliton created
potential $U$ locally increases refractive index.
According to the Fermat principle a maximum of the
refractive index serves as an attracting potential  for waves
experiencing diffraction. The diffraction of beams is analogous to the
dispersion of pulses with a spectrum in the range of
anomalous GVD.  If, however, GVD is normal,
then the refractive index maxima serve as  repelling potential
barriers and the index minima become attracting potentials for
dispersing pulses. Therefore to explain the trapping of  radiation
at the short-wavelength edge of the continuum, we should look for a minimum
in the refractive index. An alternative, but equivalent, view of the problem  is to note
the similarity between the right-hand side of Eq. (\ref{cnls}) and the quantum mechanical Shr\"odinger
equation, in which  the dispersion operator $\partial_{t}^2$ corresponds
to  kinetic energy and the function $U$ to  potential energy.
Dispersion at the short-wavelength edge of the continuum is normal,
i.e., $d<0$. Therefore Eq. (\ref{cnls}) coincides with the Schr\"odinger equation for
quantum particles having positive mass. Thus,  $U$ is  a repelling potential for any given $z$
(see the dashed line in Fig. 3(a)), in which case it
can not trap waves and so can not
lead to the formation of  localized non-dispersive light pulses
as observed in experiments and modeling. What this potential can do, however,
is to serve as a barrier for the incident waves. This is why the dispersive spreading
of the short-wavelength  pulse in Figs. 2(d,e) only occurs with longer time delays.
In other words, no dispersive waves can pass through the soliton.

 Let us switch now from the laboratory frame
to a noninertial frame of reference accelerating  together with the soliton,
i.e. together with the potential $U$. In order to achieve this we replace
$t$ in Eq. (\ref{cnls}) with the  variable $\tau$ (as defined above)
and take account of the fact (obtained from the XFROG data)
that the group velocities of the pulses at the long- and short-wavelength edges
 are matched, $v_s=v_l$. The possibility for this matching within the spectral range of interest is illustrated in the Fig. 1(c).
Group velocities are uniquely related to frequencies
through the dispersion law, and so we require that the phase of the
field in the new reference frame  be shifted. This is achieved by making
the substitution
$A=\psi(z,\tau)\exp\{i\tau gz/(4d)-ig^2z^3/(24d)\}$ \cite{bel}.
The amplitude $\psi$  in the non-inertial frame obeys
\begin{equation}
-i\partial_z\psi=[d\partial^2_{\tau} + U+\tilde g\tau]\psi, ~\tilde g=-g/(4d)>0\label{cnls1}
\end{equation}
Therefore the potential acting on the short-wavelength pulse in the accelerating
frame has a minimum at $\tau=\tau_0$, see the full line in Fig.
3(a). The walls of the new attracting potential, $(U+\tilde g\tau)$,
are formed, on one side, by the refractive index change created by the
soliton through the Kerr and Raman nonlinearities and, on the other
side, by the linear growth of the refractive index existing in the
accelerating frame of reference. The linear part of the potential is
equivalent in its essence to the effective gravity emulated in the
accelerating reference frames and exerts a
force on photons stopping them from dispersing in the direction of
increasing $\tau$. To describe the situation qualitatively it is
instructive to appeal to the example of an elevator loaded with balls,
see Fig. 4. If the elevator moves with a constant
velocity in outer space, then the balls are weightless, they
randomly kick the walls and diffuse upwards, while the elevator floor
prohibits the diffusion downwards. This corresponds to
the  photons freely dispersing behind the soliton when the Raman
effect is switched off, see Figs. 2(d,e). However, when the elevator is
pulled with an acceleration, the balls  pile up on the floor,
just as if someone switched on gravity.  This corresponds to the
process of the radiation trapping  with the soliton acting as
the elevator floor. An analogy between the full
potential $U+\tilde g\tau$ and the potential felt by cold atoms
placed close to an atomic mirror and subjected to gravity
should be mentioned also \cite{atom}.

\begin{figure}
\includegraphics[width=0.5\textwidth]{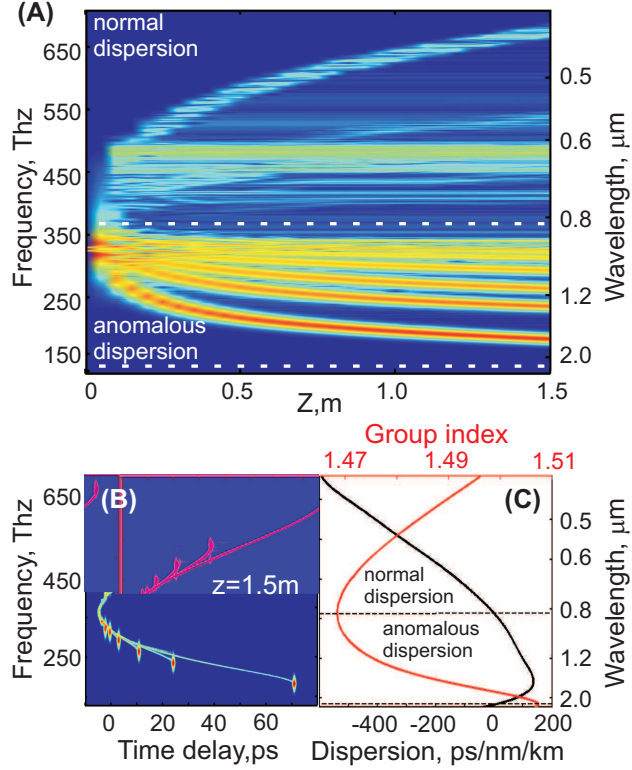}
\caption{{\bf Figure 1. Numerical simulation of supercontinuum generation in a  photonic
crystal fiber pumped with $200$fs pulses at 850nm and having
$6$kW peak power.} (A) Spectral evolution along the fiber length. (B)
Time-frequency resolved signal for $z=1.5$m. The localised pulses in the
anomalous and normal dispersion ranges correspond to the solitons and
to the trapped radiation, respectively.
(C) Group index, i.e. speed of light
divided by the group velocity, (red line) and dispersion $D$ (black
line) typical for  photonic crystal fibers used in supercontinuum
experiments \cite{gorbach}.}
\label{fig1}
\end{figure}

We solved the eigenvalue problem for the potential $U+\tilde g\tau$ numerically
using a finite difference approach to determine all of its eigenmodes.
We have found that the strongest soliton on the long-wavelength edge
of the supercontinuum spectrum in Fig. 1(a)
traps around 20 modes on the short-wavelength edge, see Figs. 3(b,c)
for three examples. The potential barrier on the soliton side is high, but still finite,
so that some light  penetrates through the soliton barrier
creating oscillating tails. The latter are especially noticeable in
the higher order modes, see  Fig. 3(c). The  radiation  captured  by the soliton
can be represented as a superposition of these modes. Adiabatic transformation of the soliton power
and width with propagation, caused by the increasing dispersion,
induces weak adiabatic evolution of the mode parameters,
but apart from this the modes are  stationary solutions
and hence their dispersion and spectral broadening
due to XPM are suppressed. According to our approach, the
phase factor $e^{-i \tilde g z\tau}$ used in the transformation
to the accelerating frame and the evolving phases of the complex mode amplitudes
should determine the spectral dynamics  at the short wavelength of the  supercontinuum.
In order to verify this, we took the data from Fig.~\ref{fig1}
for $z=1.3m$ and represented the field at the short-wavelength of the continuum as a
superposition  of the  first 10 modes (the rest can be neglected).
The evolution of the trapped radiation over the remaining
length of the fiber is then simply calculated by applying the appropriate
phase shifts to the individual modes. The spectrum of the resulting field
is again compared  with the data from Fig.~\ref{fig1}.
Fig.~\ref{fig5} demonstrates a good agreement between the two approaches  in predicting both
the position of the short-wavelength edge of the continuum and its spectral shape.
Small discrepancies in the shape of the spectral peaks get more noticeable
for larger $z$ and are due to the adiabatic changes of the soliton
parameters, which are not accounted for in Eq.~(\ref{cnls1}).

Summarizing: we have explained  the physical mechanisms
behind existence of the non-dispersive and continuously blue shifting
localized states of light at the short-wavelength edge
of an expanding  supercontinuum spectrum in optical fibers.
This effect  has been observed in many recent experiments with photonic crystal
fibers pumped by femtosecond pulses in the proximity of the zero dispersion point.
We have demonstrated that the light at the short-wavelength edge of the continuum
is trapped, on one side,  by  nonlinear refractive index variations induced
by  Raman solitons existing on the long-wavelength edge
and, on the other side, by a linear  gravity-like
force originating from the fact that the soliton moves with acceleration.
Our findings are an important addition to the
list of  physical effects known to be capable of suppressing  dispersive spreading of  short pulses
and of transforming their frequency \cite{agrawal}, which  open new research avenues
in the areas of fiber based frequency conversion and optical solitons.

{\bf Acknowledgement}\\
This work has been supported by EPSRC.

{\bf Competing financial interests}\\
The authors declare that they have no competing financial interests.

\newpage

\begin{figure}
\includegraphics[width=0.5\textwidth]{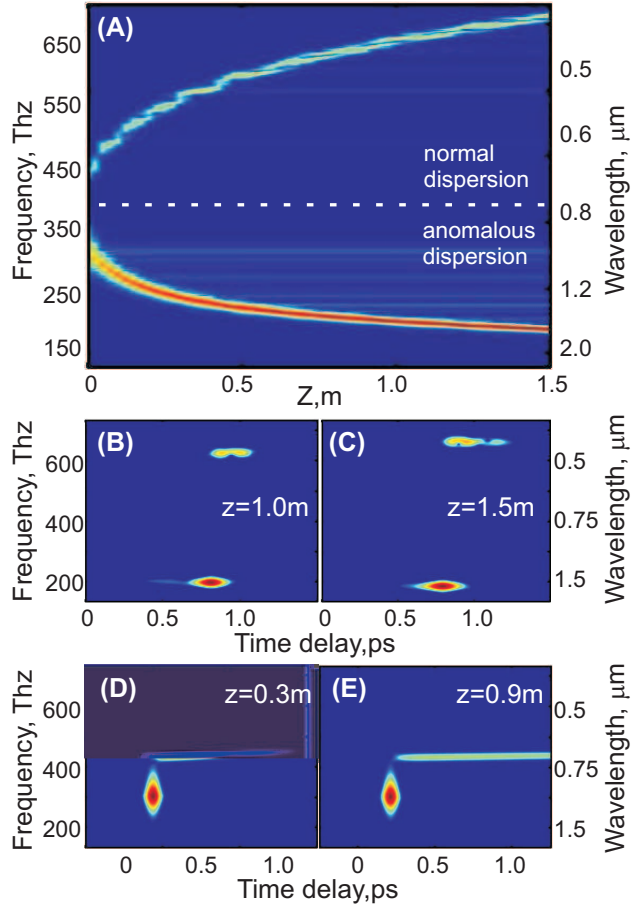}
\caption{{\bf Figure 2. Radiation  trapping  by a soliton.}
(A) Spectral evolution along the fiber length showing that the frequency of the
short-wavelength radiation is continuously drifting further to the blue,
simultaneously with the continuous red shift of the soliton.
(B,C) Time frequency resolved signal for $z=1$m and $1.5$m showing that the blue
shifting pulse does not disperse with propagation
and is delayed with respect to the soliton.
(D,E) The same as (B,C), but with the Raman effect switched off. One can see dispersive
 spreading of the short-wavelength pulse.
Parameters of the input pulses:  the soliton width and peak power are
$10$fs and  $38$kW (as found from Figs. 1(B)), the short-wavelength
pulse width and peak power are $30$fs and $2$kW.
Time delay between the pulses is $50$fs.}
\label{fig2}
\end{figure}

\begin{figure}
\includegraphics[width=0.5\textwidth]{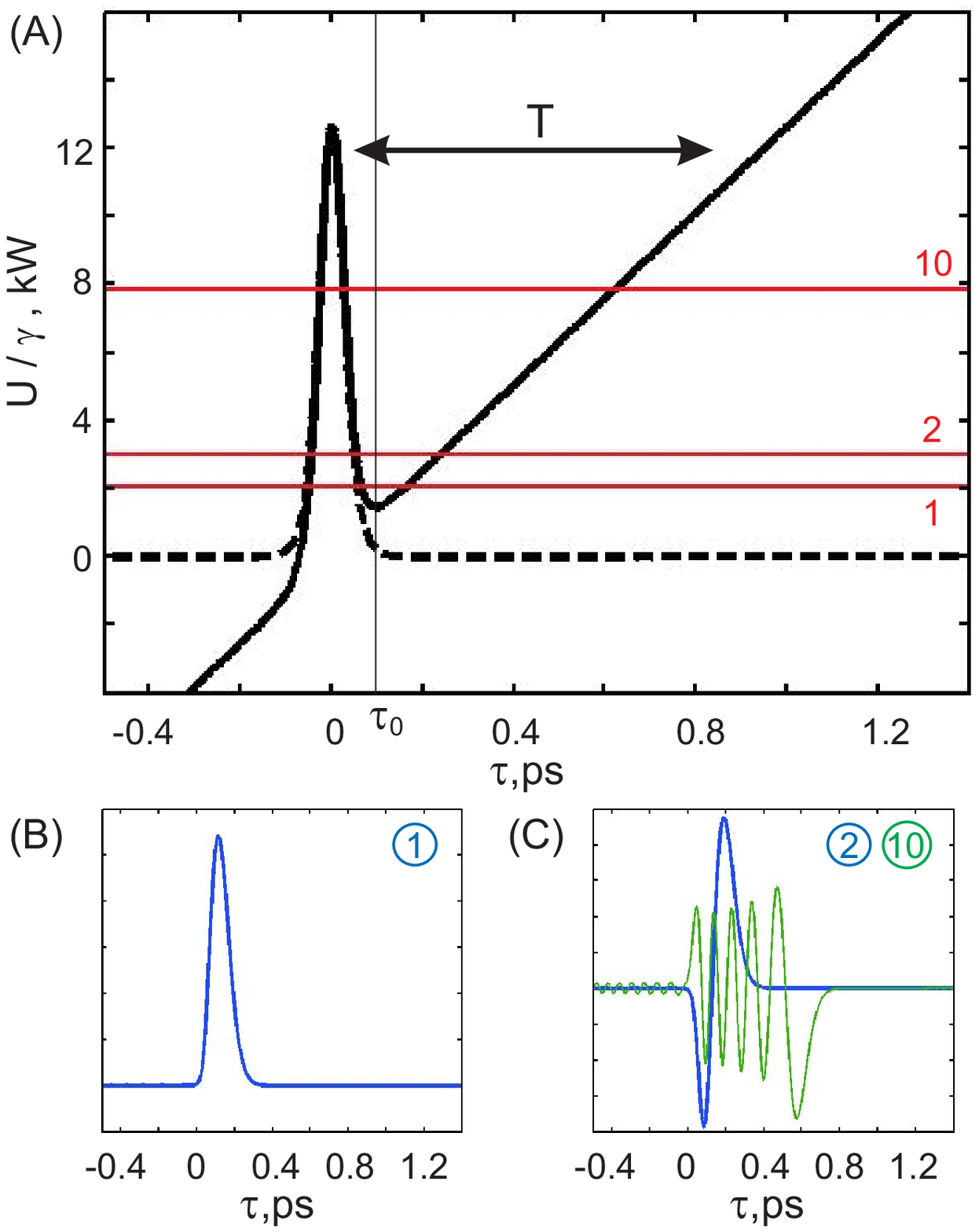}
\caption{
{\bf Figure 3. Effective potential and quasi-trapped states.}
(A) Dashed line shows the potential $U$.
Full line shows the potential $U+\tilde g\tau$,
which takes account of the gravity-like force.
Red horizontal lines correspond to the
effective 'energy levels' of the 1st, 2nd and
10th modes  shown in (B) and (C).}
\label{fig3}
\end{figure}

\begin{figure}
\includegraphics[width=0.5\textwidth]{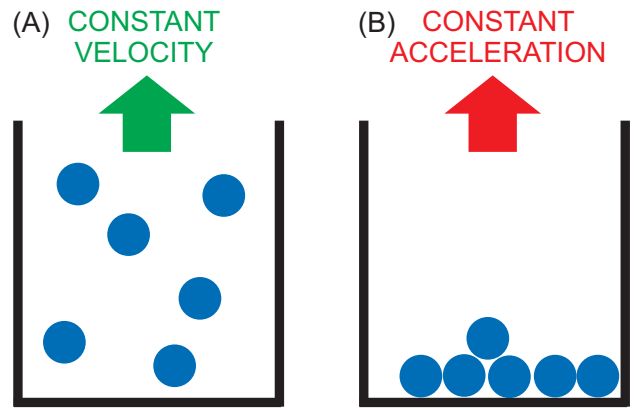}
\caption{
{\bf Figure 4.} Dispersion of the blue radiation in the laboratory frame, as
in Figs. 2(D,E),
is analogous to weightless balls flying around an elevator
moving with a constant velocity, see (A).
Trapped states of the blue radiation as in Figs. 2(B,C) and 3(B-E) are
analogous to the balls piled on the floor of the elevator
moving with a constant acceleration.}
\label{fig4}
\end{figure}

\begin{figure}
\includegraphics[width=0.5\textwidth]{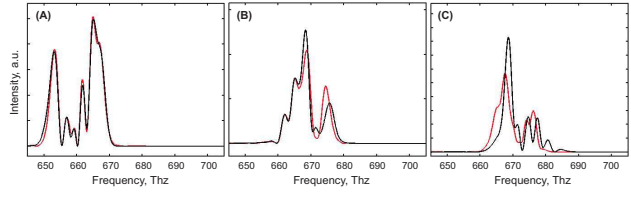}
\caption{
{\bf Figure 5. The field at the short-wavelength edge
of  the supercontinuum can be represented as
a superposition of the modes of the potential
induced by the accelerating soliton, see Fig. 3.}
Spectral peaks at the blue edge of the supercontinuum
as in Fig.~\ref{fig1} at different propagation distances:
(A) $z=1.3$m; (B) $z=1.4$m; (C) $z=1.5$m. Black and red lines correspond to the
data from Fig.~\ref{fig1} and to the results obtained
using expansion over 10 modes, respectively.}
\label{fig5}
\end{figure}



\end{document}